# Ferroelectricity with Asymmetric Hysteresis in Metallic LiOsO$_3$ Ultrathin Films

Jinlian Lu[1,2†], Gong Chen[3†], Wei Luo[1,2], Jorge Íñiguez[4,5*], Laurent Bellaiche[6*], Hongjun Xiang[1,2*]

[1]*Key Laboratory of Computational Physical Sciences (Ministry of Education), State Key Laboratory of Surface Physics, and Department of Physics, Fudan University, Shanghai 200433, China*

[2]*Collaborative Innovation Center of Advanced Microstructures, Nanjing 210093, China*

[3]*Department of Physics, University of Science and Technology of China, Hefei 230026, China*

[4]*Materials Research and Technology Department, Luxembourg Institute of Science and Technology, 5 Avenue des Hauts-Fourneaux, L-4362, Esch/Alzette, Luxembourg*

[5]*Physics and Materials Science Research Unit, University of Luxembourg, 41 Rue du Brill, L-4422 Belvaux, Luxembourg*

[6]*Physics Department and Institute for Nanoscience and Engineering University of Arkansas, Fayetteville, Arkansas 72701, USA*

†J. L. and G. C. contributed equally to this work.

*E-mail: jorge.iniguez@list.lu, laurent@uark.edu, hxiang@fudan.edu.cn

**Abstract**

Bulk LiOsO$_3$ was experimentally identified as a "ferroelectric" metal where polar distortions coexist with metallicity [*Shi et al., Nature Materials 12, 1024 (2013)*]. It is generally believed that polar displacements in a "ferroelectric" metal cannot be switched by an external electric field. Here, via comprehensive density functional theory calculations, we demonstrate that a two-unit-cell-thick LiOsO$_3$ thin film exhibits a ferroelectric ground state having an *out-of-plane* electric dipole moment that can be switched by an external electric field. Moreover, its dipole moment-versus-electric-field hysteresis loop is asymmetric because only surface Li ions' displacements are reversed by external electric field whereas the field-induced force on inner Li atoms are nearly fully screened by itinerant electrons. As a relevant by-product of our study, we also extend the concept of "Born effective charge" to finite metallic systems, and show its usefulness to rationalize the observed effects.

**Main text**

Ferroelectricity is known as a property of an insulator having a spontaneous electric polarization whose direction can be reversed by an electric field. About 50 years ago, Anderson and Blount [1] proposed an interesting concept of "ferroelectric (FE) metal". Recently, Shi *et al.* experimentally demonstrated that metallic bulk $LiOsO_3$ undergo a "FE-like" phase transition from a high-temperature non-polarized paraelectric state to a low-temperature "FE" polarized state [2] as a result of zero center phonon softening [3-5]. The combination of metallicity and non-centrosymmetric structures may produce unique physical properties, such as non-centrosymmetric superconductivity [6-9], unconventional optical responses [10, 11], magnetoelectricity [12-14], highly anisotropic thermopower response [15], and FE photovoltaic properties [16]. However, the so-called bulk "FE metal" usually does not display ferroelectricity since the polarization in the "FE metal" cannot be switched by an electric field due to the screening of the itinerant electrons. In fact, any "FE metal" is also referred to as a "polar metal" in the literature [17].

In order to further expand the applications of "FE metals", it is however highly desirable to switch the polarization in a "FE-metal" with an electric field. In 2014, Xiang proposed that the polarization may be controlled by the electric field if the current flow perpendicular to the slab is blocked [18]. Similarly, Filippetti *et al.* [19] and Luo *et al.* [20] predicted the possibility of a polarization reversal in $Bi_5Ti_5O_{17}$ layers and two-dimensional (2D) CrN, respectively. Very recently, Fei *et al.* experimentally observed such polarization reversal in a 2D $WTe_2$ FE metal by applying an electric field [21]. Despite these remarkable progresses, how and why the polarization in a "FE metal" is switched by an external electric field is not clear.

In this Letter, we systematically investigate the ferroelectricity of metallic polar $LiOsO_3$ films with various thicknesses [see Sec. I and II of Supplementary Material (SM) [22]]. We search the ground state configuration and tune the spontaneous electric dipole moment (here, we used the spontaneous electric dipole moment rather than the common word "polarization" for describing polar intensity in metallic films more accurately) by an external electric field. For the 2-unit cell-thick (2-UC) film, the ground state is FE, and its net electric dipole moment can be asymmetrically switched and reversed by an external electric field, due to an "only" partial screening of bound charges via itinerant electrons. For thicker films, even-UC films exhibit a "ferrielectric" (FI) ground state while odd-UC films adopt an "antiferroelectric" (AFE) ground state, but none of them exhibits a field-induced switchable net electric dipole moment -- because of high energy barriers.

**Structures and dipole moments of thin films.** Bulk LiOsO$_3$ belongs to the centrosymmetric R$\bar{3}$c space group at room temperature. In this phase, Li and O ions lie at an identical plane (marked as the O$_3$ plane, which is perpendicular to pseudo-cubic [111] direction as indicated via blue dot-line in Fig. 1) and Os ions are exactly halfway between two successive O$_3$ planes. As a result, there is no net electric dipole moment. Below 140 K, bulk LiOsO$_3$ transforms into a non-centrosymmetric R3c structure [2], which exhibits Li ionic motions along the pseudo-cubic [111] direction – therefore generating a spontaneous electric dipole moment.

LiOsO$_3$ films are then constructed from the bulk structure and with the out-of-plane direction being along the polar [111] direction. By comparing the formation energies ($U_{form}$) of different surface terminations (see Fig. S1 of SM [22]), the case with the lowest formation energy (i.e., LiO$_3$-LiO$_3$) is chosen as the surface terminations of films. LiOsO$_3$ films with thicknesses ranging from 1-UC to 9-UC are investigated. We consider films with equivalent top and bottom surfaces; in this way, we have no built-in electric field that may favor one polar orientation. The 1-UC film is not included in our results since too low ratio of cations in 1-UC film leads to an unreasonable valence (+10 for the Os ion) and high formation energies.

There are three possible relative locations of each Li ion with respect to the O$_3$ plane (i.e., above, on or below the plane). To determine the ground state, all combinations of the locations of Li ions are carefully examined. Figure 1 show the fully relaxed configurations and also reports the corresponding energies for the stable combinations of 2-UC, 3-UC and 4-UC films. It is found that the configurations with the Li ion located on the $O_3$ plane (e.g., $II_e$ or $IV_b$ structure) are generally higher in energy than those with shifted Li ions (i.e., for which Li ions is above or below the $O_3$ plane), due to the off-centering (polar) instability associated to the Li ions [3, 23]. For the 2-UC films, the $II_a$ FE configuration is the most energetically favorable, in which all Li ions shift towards the same direction with respect to the $O_3$ planes like in bulk. Our test calculations also show that the $II_a$ FE configuration is also more stable than in-plane AFE structures (not shown here). The computed phonon spectrum [24, 25] (see Fig. S2 of SM [22]) indicates that this FE state is dynamically stable. The computed electric dipole moment of this FE state reaches $60\times10^{-3}$ e·Å/Li (see Fig. 1).

For the n-UC (n>2) films, symmetric or quasi-symmetric configurations are preferred, with dipoles being in opposite directions between the top and bottom surfaces to decrease the depolarization field and thus the energy. The odd-UC films with an even number of Li ions are thus AFE, while the even-UC films with an odd number of Li ions are FI due to the shifted central Li ions. As shown in Fig. 1, the obtained AFE solutions ($III_a$, $IV_b$) are peculiar ones, as they involve a longitudinal modulation of the polar distortion and thus resemble tail-to-tail charged ferroelectric domain walls. As indicated in Fig. 2(a), the net electric dipole moment of all FI states is extremely small, which will be explained shortly.

**"Born effective charges" in metallic thin films.** The Born effective charge (BEC) [26] is a useful physical quantity to describe the response of an insulator under an electric field. The BEC tensors can be computed as $Z^*_{k,\beta\alpha} = \Omega \frac{\partial P_\beta}{\partial u_{k\alpha}} = \frac{\partial F_{k\alpha}}{\partial E_\beta}$, suggesting that they can be calculated as the change of the polarization with respect to the atomic displacement or as the additional atomic force due to the presence of an electric field. Although the BEC is originally defined for insulators, here we generalize for the first time the usage of the BEC to metallic systems. As we will show shortly, "BEC" is also useful for describing the electric field response in metallic thin films. Practically, we compute the "BEC" of paraelectric thin films that are "cut" from the bulk paraelectric $R\bar{3}c$ structure. For each Li ion of a given thin film, we move the Li ion along the c-axis (i.e., perpendicular to the plane of the thin film) by a small amount (e.g., 0.2 Å), and then extract the "BEC" via $Z^*_{k,\beta\alpha} = \Omega \frac{\partial P_\beta}{\partial u_{k\alpha}}$.

For an insulator, the polarization can be computed with the Berry phase approach [27]. However, the Berry phase theory of polarization is not applicable to a metal. Fortunately, we can compute the dipole moment by a direct integration of the charge density of a metallic thin film. Our results [see Fig. 2(b)] show that the surface Li ion has a "BEC" of about 0.3 e, while the "BEC" of the internal Li ion is much smaller (less than 0.03). From 2-UC to 4-UC thin films, the "BEC" of the internal Li ion becomes closer and closer to zero. In the 4-UC case, the subsurface Li ion also has a very small "BEC". These results suggest that the internal Li ions are screened by the itinerant electrons, while the screening of the surface Li ions is not complete. Thanks to the computed "BECs", we can explain the electric dipole moments of $LiOsO_3$ thin films. The magnitude of the dipole moment of 2-UC $II_c$ is only slightly smaller than that of the 2-UC $II_a$ structure (see Fig. 1) because the change of the magnitude of the dipole moment is mainly due to the displacement of the internal Li ion and such internal Li ion has a very small "BEC". The small "BEC" of the internal Li ions also results in small electric dipole moments for 2-UC $II_b$, 2-UC $II_d$, 4-UC $IV_a$, 6-UC and 8-UC thin films [see Fig. 1 and Fig. 2(a)]. We note that the behavior of insulating films under electric field could be rather different from that of metallic films (see Sec. III of SM [22]).

**Net electric dipole moment reversed by an applied electric field.** By definition, for a material to be FE, the spontaneous polarization (or electric dipole moment) must be reversible. We thus applied an external electric field antiparallel to the polar direction on all films, to try to reverse the electric dipole moments. Figure 3(a) shows the electrical-dipole moment-versus-electric-field hysteresis loop of the 2-UC film, starting from the $II_a$ FE configuration. A first phase transition point appears at the electric field of $5\times10^9$ V/m. At this point, the motion of the Li-down ions along the out-of-plane $c$-axis [Li-up, Li-middle and Li-down are indicated in Fig. 4(a)] is reversed (i.e., the Li-down ion is now above, rather than below, the $O_3$ plane), and the film now adopts the $II_b$ configuration (see Fig. 1). As the electric field is further increased to $8\times10^9$ V/m, the motion of Li-up along the *c*-axis is also reversed, and the film now possesses the II$_c$ configuration. However, the motion of the Li-middle ion along the *c*-axis never switches, even when the electric field reaches $2.5\times10^{10}$ V/m, which implies that a reversed state that would be symmetry-equivalent to the II$_a$ FE configuration cannot be achieved. Then, we reduce the electric field gradually down to 0 V/m. During that decrease, the system remains in the asymmetric II$_c$ configuration, but its net dipole moment decreases down to a zero-field value that is smaller than that of the II$_a$ FE configuration -- which is due to the fact that the central Li ions never revert their displacements along the *c*-axis. When the electric field is further reduced to $-2\times10^8$ V/m, the II$_d$ configuration with restored Li-up (i.e., the Li ion shifts back to be below the O$_3$ plane) is observed. Once the electric field drops to $-8\times10^8$ V/m, both Li-up and Li-down are restored to their original positions, so that the configuration returns to II$_a$. As the electric field is decreased further, the II$_a$ configuration is retained and the corresponding electric dipole moment linearly increases in magnitude. This hysteresis loop is therefore asymmetric since the Li middle is not shifted, which differs from the case of ordinary FE materials.

Note also that one usually sandwiches insulating FE materials with metallic electrodes realize switching of ferroelectricity. However, in the "FE metal" case, if one contacts metallic film with electrodes, there will be a current flow and no electric field will be effectively applied. In practice, to apply electric fields to metallic films, one needs to grow some buffer insulating layers [18, 19], or straightly bring electrodes/charges to the proximity of a free-standing film [18]. As a reference, in our calculation, the dipole layer that generates the external fields considered here ($8\times10^9$ V/m) has a surface dipole moment density of $3.38\times10^{-11}$ C/m, which amounts to a polarization charge of $7.08\times10^{-2}$ C/m$^2$. In our calculations, the electric field reaches values up to $8\times10^9$ V/m, which appears to be high. However, theoretical *ab-initio* electric fields are typically larger by one or two order of magnitudes than experimental ones [28, 29]. Moreover, a recent measurement showed that electric fields as high as $\simeq 3\times10^8$ V/m can be achieved in some multiferroic films [30]. As a result, theoretical fields of the order of $10^{10}$ V/m are, in fact, reasonable, since they should correspond to experimental fields of the order of about $10^8 - 10^9$ V/m.

The net electric dipole moment for the 4-UC film is also plotted as a function of electric field in Fig. 3(b). If we start from the IV$_a$ ferrielectric ground state with a negative electric dipole moment [see Fig. 2(a)], a positive electric field only results in a gradual change of the electric dipole moment towards the positive *c*-direction, because all Li ions have displacements that cannot be reversed in direction and only Li-up ions shifts slightly their motions under the electric field. As a result, there is no electrical hysteresis loop. Moreover, after the removal of the electric field, the 4-UC film always returns to its FI ground state. Similarly, a negative electric field only results in an enhancement of the magnitude of the (negative) electric dipole moment without inducing a phase transition. Therefore, the behavior of the 4-UC film under electric field is dramatically different from that of the 2-UC film.

Let us now understand why there is ferroelectricity in 2-UC LiOsO$_3$ film, but not in the 4-UC case. For that, we computed the field-induced forces and the reversal barriers of the Li ions for 2-UC FE and 4-UC FI films (see Fig. 4). The field-induced force on the inner Li ions for the 2-UC and 4-UC films nearly fully vanish even when electric field reaches up to $2\times10^{10}$ V/m. In contrast, the surface Li ions (Li-up and Li-down) undergo a notable field-induced force. This result indicates that the electric field inside the films is nearly completely screened by itinerant electrons, in agreement with the "BEC" results, and all inner Li ions are thus pinned.

Since surface Li ions in the 2-UC and 4-UC films are both experiencing significant forces under the electric field, it remains unclear why the electric dipole moment in the 4-UC film cannot be switched by an electric field. To understand the difference between 2-UC film and 4-UC film, we calculated the different reversal barriers of the Li ions of 2-UC and 4-UC films. In the 2-UC thin films, the barrier is relatively low, so that the surface Li ions can be reversed by the electric field. On the contrary, in the 4-UC films, the barrier for the surface Li-down ion is extremely high; therefore the surface Li-down ion can only move slightly, but it can never crossed the O$_3$ plane. This may be because, in the case of the 4-UC film, we have more free carriers than the 2-UC film (see Fig. S3 and Table S1 of SM [22]). Note also that for the 4-UC film, the Li-up ions do not cross the $O_3$-plane during the structural path associated with Fig. 4(a), which explains why their barrier is null in Fig. 4(b).

In conclusion, by applying an electric field, we theoretically demonstrate the existence of ferroelectricity in ultrathin $LiOsO_3$ films (of a thickness of about 2-UC). A resulting asymmetrical hysteresis loop is achieved due to pinned middle Li ions. Excitingly, Fei *et al.* [21] experimentally and recently observed such polarization reversal in a 2D $WTe_2$ FE metallic film by applying an electric field, which is consistent with our theoretical results on metallic $LiOsO_3$ thin films. Moreover, when the film thickness is increased, no switchable ferroelectricity is found, which implies that films having minimal thicknesses are required for switching polar displacements in metal and to make miniaturized devices.

**ACKNOWLEDGMENTS**

Work at Fudan is supported by NSFC (11374056), the Special Funds for Major State Basic Research (2015CB921700), Program for Professor of Special Appointment (Eastern Scholar), Qing Nian Ba Jian Program, and Fok Ying Tung Education Foundation. H. X. also thank the support of the Department of Energy, Office of Basic Energy Sciences, under Award No. DE-SC0002220. L. B. thanks the Air Force Office of Scientific Research under Grant No. FA9550-16-1-0065. We also acknowledge funding from the Luxembourg National Research Fund through the inter-mobility (Grant No. 15/9890527 GREENOX, J. I. and L. B.) and Pearl (Grant No. P12/4853155 COFERMAT, J. I.) programs. Some computations were also made possible owing to MRI Grant No. 0722625 from NSF, ONR Grant No. N00014-15-1-2881 (DURIP), and a Challenge grant from the Department of Defense.

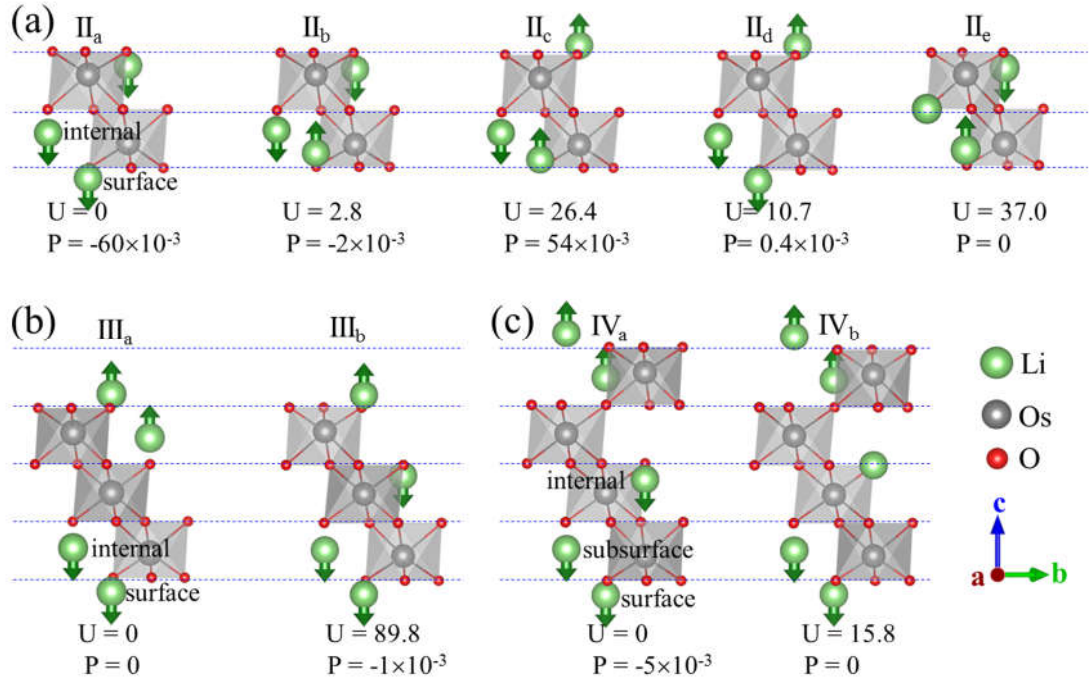

FIG. 1. Side views of metastable states and ground state of (a) 2-UC, (b) 3-UC and (c) 4-UC LiOsO$_3$ films. The values of the relative energy (U, meV/Li) and the electric dipole moment value (P, e·Å/Li) are also provided here.

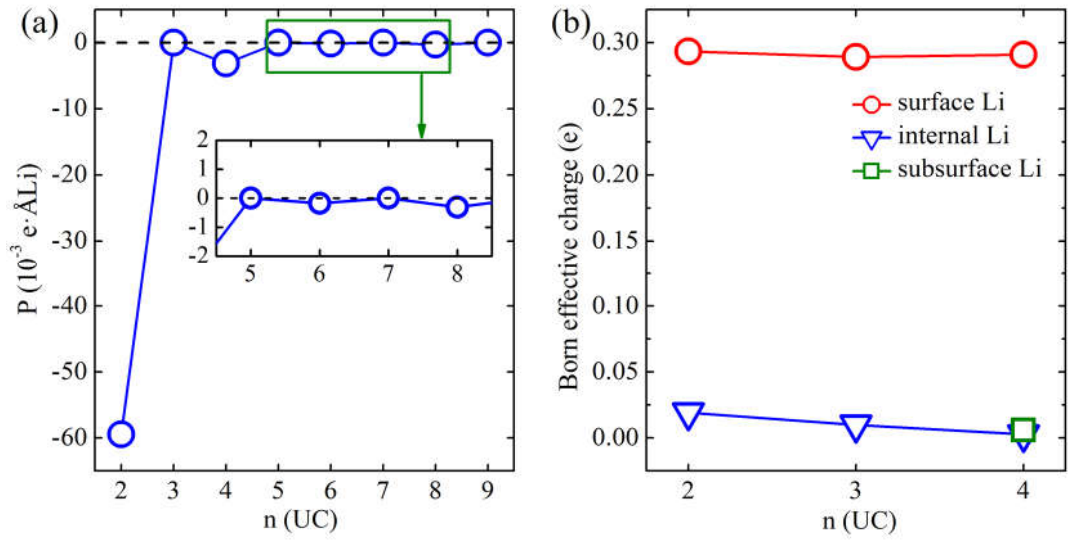

FIG. 2. (a) The electric dipole moment (P) of the lowest energy films as a function of thickness (n). (b) Born effective charge of Li ions of 2, 3 and 4-UC $LiOsO_3$ films. The surface, subsurface, and internal Li ions are indicated in Figure 1.

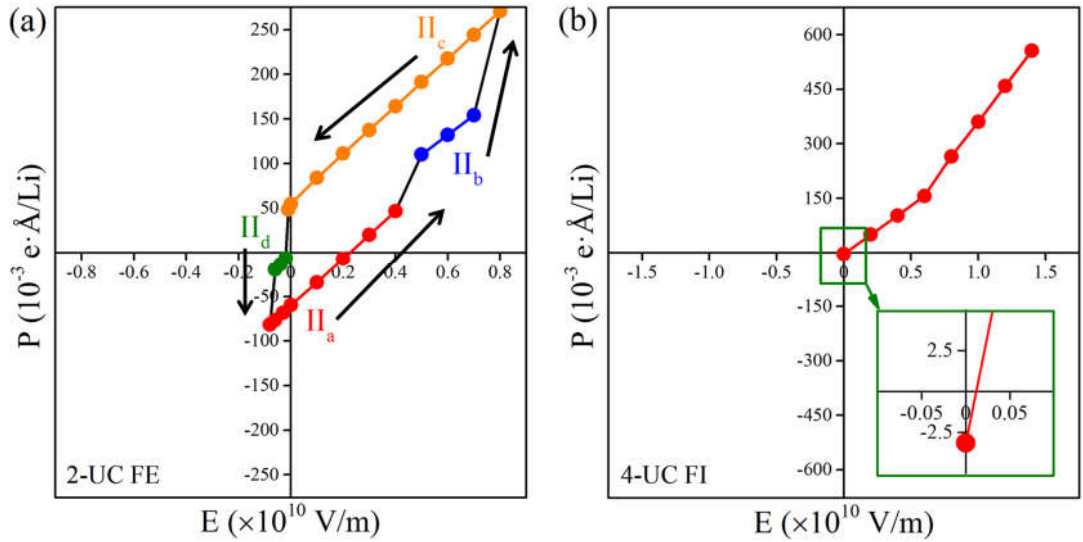

FIG. 3. (a) The asymmetrical electric-dipole-moment-*versus*-electric-field hysteresis loop of 2-UC FE film. The Red, blue, orange and green circles correspond to the four different $II_a$, $II_b$, $II_c$ and $II_d$ structures (see Fig. 1). The arrow indicates the direction of change of the electric field. (b) Dependence of electric dipole moment of 4-UC FI film on the external electric field. The inset is an enlargement around the zero field.

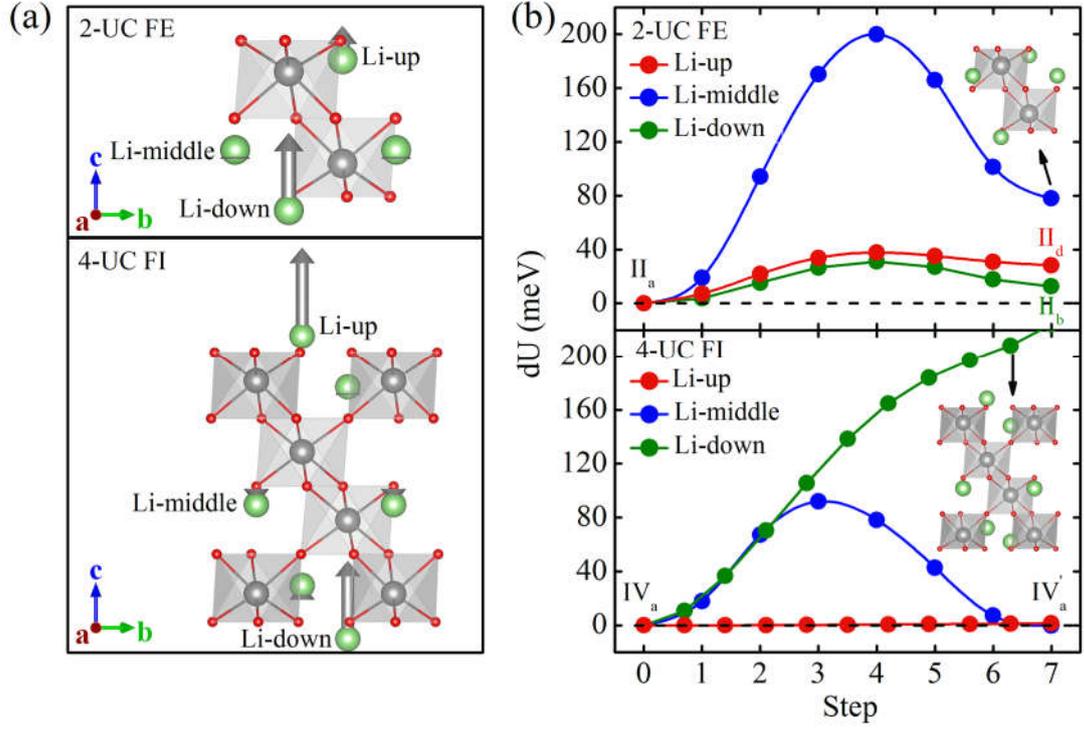

FIG. 4. (a) Atomic forces (gray arrows) of Li atoms induced by electric field of $5\times10^9$ V/m for 2 and 4-UC films. (b) The energy barrier (dU) of Li atoms in different positions [indicated in (a)] in the electric dipole moment switching process. $IV_a'$ is symmetry-equivalent state to the $IV_a$ FI configuration.